\documentclass{pasj00}
\draft

\begin{document}
\SetRunningHead{S. Kato}
{Quasi-periodic oscillations on  
spin-induced deformed disks}
\Received{2005/12/10}
\Accepted{2005/05/11}

\title{
Quasi-Periodic Oscillations Resonantly Induced on Spin-Induced
Deformed-Disks of Neutron Stars
}

\author{Shoji \textsc{Kato}%
  }
\affil{Department of Informatics, Nara Sangyo University, Ikoma-gun,
  Nara 636-8503}
\email{kato@io.nara-su.ac.jp; kato@kusastro.kyoto-u.ac.jp}


%

\KeyWords{accretion, accretion disks 
     --- kHz quasi-periodic oscillations 
     --- neutron stars 
     --- relativity 
     --- resonances 
     --- X-ray binaries 
} 

\maketitle

\begin{abstract}
We consider the situation where a spin-induced wavy perturbation is
present on disks of neutron-star low-mass X-ray binaries.
By non-linearly coupling with the perturbation,
some modes of disk oscillations resonantly interact with the 
disks. 
We examine the resonant conditions under the assumption that the disks 
are vertically isothermal.
The results show that when the spin frequency has some particular
values, resonant coupling occurs at some radii.
The modes and frequencies of these resonant oscillations are summarized.
Some of these resonant oscillations seem to explain the frequencies of
the observed kHz QPOs in objects of which the spin frequency is known.

\end{abstract}

\section{Introduction}

In many neutron-star and black-hole X-ray binaries, various kinds of
quasi-periodic oscillations (QPOs) have been observed
(see van der Klis 2004 for a recent review).
Their origin is still being debated, but many authors suspect that
they are disk phenomena in a strong graviatational field of general
relativity (e.g., Abramowicz, Klu{\' z}niak 2001;  
Klu{\' z}niak, Abramowicz 2001).

Observations of neutron-star X-ray binaries of which the
spin rate of the central star is observed show that the frequency
difference of a pair of kHz QPOs is close to, or half of, the neutron-star 
spin frequency.
Regarding this frequency relation as the essence of kHz QPO phenomena in 
neutron-star X-ray binaries, Lamb and Miller (2003) proposed a 
sonic-point spin-resonance model.
On the other hand, extending their model of 3:2 resonances of 
high-frequency QPOs in black-hole X-ray binaries, Klu{\'z}niak et al.
(2004) and Lee et al. (2004) proposed models in which the disk responds
to external perturbations at its own eigen-frequencies
related to the radial and vertical epicyclic frequencies at a certain radius.

In a series of papers (Kato 2003, 2004a,b, 2005a,b), we proposed the idea that
the cause of kHz QPOs are related to disk deformation by a warp.
In this model, the presence of disk deformation by a warp is the
cause of kHz QPOs, leading to
resonant interactions between disk oscillations and the deformed
disks at some particular radii.
The resonance conditions due to these resonant 
processes are examined in detail (Kato 2004a, 2005a,b).
Whether the oscillations are really spontaneously excited by the resonances 
is also examined (Kato 2004b).
Based on these studies, we proposed a unified model of QPOs (Kato 2005a,b).
That is, kHz QPOs in neutron-star X-ray binaries and high-frequency
QPOs in black-hole X-ray binaries come from the same mechanism and
are explained as resonant oscillations in warped disks.

In the above-mentioned resonance model in warped disks, the spin of 
the central neutron star has no part.
In real disks, however, the presence of spin-induced perturbations is
expected on disks in addition to warps.
A disk deformed by a spin-induced perturbation is  
similar to the disk deformed by a warp in our present problem.
Hence, we can expect resonant interactions between disk oscillations and 
spin-induced deformed disks by the same mechanism as
in the case of warped disks.
In this paper, we examine this issue.
The main interest is to derive relations between
the frequencies of the resonant oscillations and the spin frequency, 
and examine whether there are cases where the frequency difference
of the resonant oscillations is close to the spin frequency, or half of it.
Finally, by applying our present model, we estimate the masses of X-ray 
sources whose spins as well as QPOs are known.

\section{Spin-Induced Oscillations by the Star--Disk Interaction}

Let us assume a non-axisymmetric perturbation (such as a burst) on
the surface of a spinning neutron-star.
Since the central star and the surrounding disk are dynamically coupled by 
a magnetic field, or radiative processes, a perturbation rotating
with the same spin frequency as the central star, say $\omega_{\rm s}$,
is imposed on the disk with various azimuthal wavenumber, $m_{\rm s}$,
and vertical wavenumber, $n_{\rm s}$.
Except for the inner transition zone of the disk, and except for the initial
epoch of the perturbations, eigen-mode oscillations with
($\omega_{\rm s}$, $m_{\rm s}$, $n_{\rm s}$) remain on the disk.
Without loss of any generality, we hereafter consider a particular set of 
$m_{\rm s}$ and $n_{\rm s}$, where $m_{\rm s}=0$, 1, 2 ... and 
$n_{\rm s}=0$, 1, 2, .... 

The next issue that we should consider here is where the normal-mode
oscillations with $\omega_{\rm s}$,
$m_{\rm s}$, and $n_{\rm s}$ preferentially exist on disks.
To discuss this issue we consider the dispersion relation.
The dispersion relation describing small-amplitude local wavy
perturbations in geometrically thin and vertically isothermal disks is 
(e.g., Kato et al. 1998; Kato 2001)
\begin{equation}
    [(\omega-m\Omega)^2-\kappa^2][(\omega-m\Omega)^2-n\Omega_\bot^2]
      -k^2c_{\rm s}^2(\omega-m\Omega)^2= 0,
\label{1}
\end{equation}
where $\omega$, $m$, and $k$ are, respectively, frequency and azimuthal and 
radial wavenumbers of perturbations. 
Furthermore, $n$ is zero or a positive integer related to the number of nodes 
in the vertical direction.
The quantities $\Omega(r)$, $\kappa(r)$, and $\Omega_\bot(r)$ are, 
respectively,
the Keplerian frequency and the epicyclic frequencies in the radial
and vertical directions, where $r$ is the radial distance from the $z$-axis
of cylindrical coordinates ($r$, $\varphi$, $z$) whose origin is at the
center of the central object and the $z$-axis is in the direction
perpendicular to the disk plane.
The quantity $c_{\rm s}(r)$ is the isothermal acoustic speed on disks.

This dispersion relation shows that there are two modes of disk oscillations:
one is the modes whose $(\omega-m\Omega)^2$ is smaller than $\kappa^2$, and
the other is those whose $(\omega-m\Omega)^2$ is larger than
$n\Omega_\bot^2$, when $n\not= 0$.
The former is the g-modes and the latter the p-modes.
In the case of $n=0$, the oscillation modes present on the disks are only
those whose $(\omega-m\Omega)^2$ is larger than $\kappa^2$.
They are classified as p-modes.
The propagation regions of these oscillations on the $\omega$--$r$ plane are 
described, for example, by Kato et al. (1998) and Kato (2001).
The propagation region of the g-mode oscillations, for example, is
$-\kappa<\omega-m\Omega<\kappa$, for a given $\omega$ and $m$.
The outside of this region is an evanescent region, where
the amplitude of the oscillations decreases exponentially.
In the propagation region, apart from the boundaries, however,
the wavelength of the oscillations becomes quite short, since the 
difference between $(\omega-m\Omega)^2$ and $\kappa^2$ increases there, and it
must be compensated by the term with $k^2c_{\rm s}^2$ in the second term 
of the left-hand side of the dispersion relation.
In this sense, the g-mode oscillations have a large amplitude and a relatively
global structure only in the region near to the boundaries, i.e., 
around $\omega-m\Omega\sim\kappa$, or
$\omega-m\Omega\sim-\kappa$.
Similarly, the p-mode oscillations have a large amplitude and relatively 
global structure only in the region 
around the radii where $\omega-m\Omega\sim n^{1/2}\Omega_\bot$, or 
$\omega-m\Omega\sim-n^{1/2}\Omega_\bot$, when $n\not= 0$.
In the case of $n=0$, the corresponding region is around
radii of $\omega-m\Omega\sim\kappa$, or $\omega-m\Omega\sim -\kappa$.

The above consideration of the characteristics of oscillation modes
suggests that when perturbations with $\omega_{\rm s}$, $m_{\rm s}$, 
and $n_{\rm s}$ are once imposed on disks due to a non-axisymmetric 
perturbation of the stellar surface, the disk predominantely 
responds near to the radii where
\begin{equation}
   (\omega_{\rm s}-m_{\rm s}\Omega)^2=\kappa^2
\label{2.2}
\end{equation}
or 
\begin{equation}
   (\omega_{\rm s}-m_{\rm s}\Omega)^2=n_{\rm s}\Omega_\bot^2.
\label{2.3}
\end{equation}
The former is the g-mode-type response (including the p-mode-type response
with $n_{\rm s}=0$) of the disks, while the latter
is the p-mode-type response with $n_{\rm s}\not= 0$.

\section{Disk Oscillations and Their Resonant Interaction with Deformed Disks}

\subsection{Disk Oscillations}

We consider here disk oscillations with frequency $\omega$, azimuthal 
wavenumber $m$, and $n$ characterizing the vertical node number.
Discussions made in the previous section concerning dispersion relation
(1) show that if we consider the g-mode oscillations of $n\not= 0$ or 
the p-mode oscillations of $n=0$, they preferentially exist
around the radii specified by
\begin{equation}
    (\omega-m\Omega)^2=\kappa^2.
\label{3.1}
\end{equation}
The p-mode oscillations with $\omega$, $m$, and $n(\not= 0)$, on the
other hand, exist preferentially around the radii specified by
\begin{equation}
    (\omega-m\Omega)^2=n\Omega_\bot^2.
\label{3.2}
\end{equation}
In the followings we examine in what cases those disk oscillations have 
resonant interactions with the disks deformed by spin-induced perturbations.

\begin{figure}
  \begin{center}
    \FigureFile(80mm,80mm){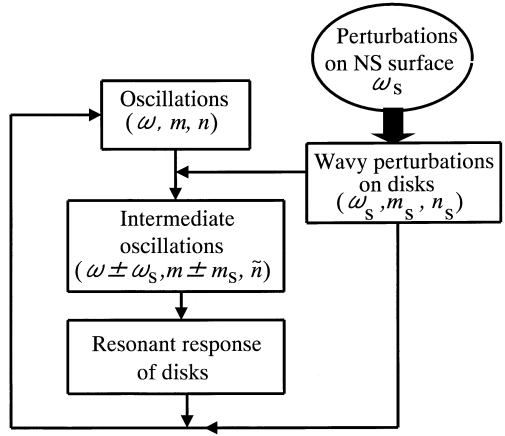}
  \end{center}
  \caption{
Feedback processes of nonlinear resonant interactions acting on
oscillations.
The original oscillations are characterized by $\omega$, $m$, and $n$.
A perturbation on the neutron star surface spinning with angular velocity 
$\omega_{\rm s}$ brings about a wavy perturbation on disks.
We consider wavy perurbations characterized by $\omega_{\rm s}$,
$m_{\rm s}$, and $n_{\rm s}$.
A non-linear interaction between the original oscillation and the
wavy perturbation gives rise to intermediate modes having  
oscillations of $\omega\pm\omega_{\rm s}$, $m\pm m_{\rm s}$, and 
${\tilde n}$, where ${\tilde n}$ is an integer between $n+n_{\rm s}$
and $\vert n-n_{\rm s} \vert$ with step two.
To these intermediate oscillations, the disk resonantly responds at
certain radii.
Then, the intermediate oscillations feedback to the original oscillations
after the resonance.
This feedback process amplifies or dampens
the original oscillations, since resonances are involved in the interaction 
processes. 
  }
  \label{fig:1}
\end{figure}

\subsection{Nonlinear Interactions}

Here, we consider nonlinear interactions between the disk oscillations 
described in subsection 3.1 and the spin-induced perturbations described 
in section 2.
The non-linear interaction process, which we consider here, are schematically
shown in figure 1.
The functional forms associated with the disk oscillations with $\omega$,
$m$, and $n$ are approximately characterized by 
exp$[i(\omega t-m\varphi)]{\sl H}_n(z/H)$
(see Kato 2004b), where ${\sl H}_n$ is a Hermite
polynomial of order $n$ with the argument $z/H$, $H(r)$ being the vertical
scale height of the disk.
The spin-induced perturbations, on the other hand, are characterized
by exp$[i(\omega_{\rm s}t-m_{\rm s}\varphi)]{\sl H}_{n_{\rm s}}(z/H)$.

Nonlinear interactions between the above two oscillations induce 
intermediate oscillations that
are characterized by 
\begin{equation}
    {\rm exp}[i(\omega\pm\omega_{\rm s})t-i(m\pm m_{\rm s})\varphi]
           {\sl H}_{\tilde n}(z/H),
\label{3.3}
\end{equation} 
where two types of oscillations are written in a single equation, i.e.,
the oscillations of the upper signs and those of the lower signs.
This kind of abbreviation is often used hereafter without notice.
Here, some comments on ${\sl H}_{\tilde n}(z/H)$ are necessary.
The product of ${\sl H}_n(z/H)$ and ${\sl H}_{n_{\rm s}}(z/H)$,
resulting from the nonlinear coupling between the oscillations (with
$\omega$, $m$, and $n$) and the spin-induced perturbations (with
$\omega_{\rm s}$, $m_{\rm s}$, and $n_{\rm s}$), should be expressed 
as a sum of Hermite polynomials, so that we can pick up an eigenmode
of oscillations from the nonlinear coupling terms.
The product of ${\sl H}_n(z/H)$ and ${\sl H}_{n_{\rm s}}(z/H)$ is a
polynomial of $z/H$ with the highest order term of 
$(z/H)^{n+n_{\rm s}}$.
This consideration shows that the product can be expressed as a sum 
of Hermite polynomials of the
order of $n+n_{\rm s}$, $n+n_{\rm s}-2$,... and $\vert n-n_{\rm s}\vert$.
Among these terms, we pick up a particular term, ${\tilde n}$.
That is, ${\tilde n}$ of ${\sl H}_{{\tilde n}}(z/H)$ in equation 
(\ref{3.3}) is
\begin{equation}
   {\tilde n}=n+n_{\rm s}, \quad {\rm or}\quad
              n+n_{\rm s}-2, \quad {\rm or} \quad ..., \quad {\rm or} \quad 
              \vert n-n_{\rm s}\vert.
\label{3.4}
\end{equation}

The oscillations described by equation (\ref{3.3}) are 
intermediate oscillations in the sense that  they 
nonlinearly couple again with the
spin-induced perturbations to feed back to the original
oscillations of $\omega$, $m$, and $n$ (see figure 1).
Such nonlinear feedback to the original oscillations can amplify or
dampen the original oscillations,  since in the process of this feedback  
a resonance process is involved, as we discuss below.
That is, the oscillations characterized by exp$[i(\omega\pm\omega_{\rm
s})t-i(m\pm m_{\rm s})\varphi]{\sl H}_{\tilde n}(z/H)$ have
resonant interactions with the disk at the radii where the dispersion
relation of the intermediate oscillations (i.e., oscillations with $\omega\pm
\omega_{\rm s}$, $m\pm m_{\rm s}$, and ${\tilde n}$)
is satisfied.
Since the effects of the pressure term on the dispersion relation (\ref{1}) 
is negligible, unless the wavenumbers are very large,
the resonance conditions can be approximatedly written as
\begin{equation}
    [(\omega\pm\omega_{\rm s})-(m\pm m_{\rm s})\Omega]^2 =
           \kappa^2
\label{3.5}
\end{equation}
or 
\begin{equation}
    [(\omega\pm\omega_{\rm s})-(m\pm m_{\rm s})\Omega]^2 =
           {\tilde n}\Omega_\bot^2.
\label{3.6}
\end{equation}
The former is a resonance condition due to  
the horizontal motions, while the latter is a resonance condition due to  
the vertical motions (Kato 2004b).

As a result of the resonances, the energy exchange between the original 
oscillations (of $\omega$, $m$, $n$) and the disk rotation occur, 
leading to amplification or damping of the oscillations (Kato 2004b).
Which occurs depends on the behavor of the oscillations at the resonant radii.
This is a complicated problem [see Kato (2004b) for the case of warped disks], 
and beyond the scope of this paper.
In this paper we only concentrate on where the resonances occur, and 
how much are the frequencies of the resonant oscillations.

For the resonances to have strong effects on the oscillations, 
they must occur in the region where both the disk oscillations 
(with $\omega$, $m$, and $n$) in consideration and the spin-induced 
perturbations (with $\omega_{\rm s}$, $m_{\rm s}$, and $n_{\rm s}$) have large 
amplitudes and moderately global structures.
This implies that the radius where equations (\ref{2.2}) or (\ref{2.3})
is satisfied and the radius  where equations (\ref{3.1}) or (\ref{3.2}) 
is satisfied are the same, and further must be the radius where one of the 
resonant conditions [equation (\ref{3.5}) or (\ref{3.6})] is satisfied.
That is, three relations 
[equations (\ref{2.2}) or (\ref{2.3}), 
equations (\ref{3.1}) or (\ref{3.2}), and equations (\ref{3.3}) or
(\ref{3.4})] must be satisfied simultaneously.
In principle, there are eight sets of combination.

\section{Resonant Conditions} 

The resonance conditions discussed above are separately examined in two cases
where a perturbation imposed on disks by the spin of the central star is 
a g-mode-type oscillation or a p-mode-type oscillation.

\subsection{{\rm g}-Mode-Type Spin-Induced Perturbations}

The spin-induced g-mode-type perturbations are 
described by equation ({\ref{2.2}).
This case is further decomposed into two cases.
The first one is the case where the disk oscillations on the
perturbed disks are g-modes.
The second one is the case where the disk oscillations are p-modes.

\subsubsection{{\rm g}-mode oscilations}

The g-mode oscillations are described by equation (\ref{3.1}).
There are two types of resonances.
The first is the resonances through horizontal motions.
In this case, the equation to be combined with equations (\ref{2.2})
and (\ref{3.1}) is equation (\ref{3.5}).
Some examination easily shows that the set of these three equations is 
simultaneously satisfied at the radius of $\kappa=0$.
The radius of $\kappa=0$ (say $r_{\rm c}$), however, is not a resonance
point in our present approximate treatment.
This is because the intermediate oscillations
behaves around $r_{\rm c}$ as $\propto 1/(r-r_{\rm c})^2$, not as
$\propto 1/(r-r_{\rm c})$.
That is, the point is not a pole, although it is a singular point 
(see Kato 2004b).
Consequently, there are no horizontal resonances in the case where the 
spin-induced perturbation is a g-mode type and the disk oscillations
are g-modes.

Next, we consider the vertical resonances.
In this case the resonance condition is equation (\ref{3.6}), not
equation (\ref{3.5}), and  the set of equations to be discussed are
equations (\ref{2.2}), (\ref{3.1}), and (\ref{3.6}).
Eliminating $\omega$ and $\omega_{\rm s}$ from these equations, we have
\begin{equation}
    \kappa={1\over 2}{\tilde n}^{1/2}\Omega_\bot, \quad ({\rm GGV})
\label{4.1}
\end{equation}
where the facts that ${\tilde n}\not= 0$ and $\kappa^2<
\Omega_\bot^2$ have been used, and GGV stands for the spin-induced 
perturbations are the g-{\it mode type}, the disk oscillations are 
g-{\it modes}, and the resonances are {\it vertical}.
Equation (\ref{4.1}) shows that resonances are possible only when 
${\tilde n}=1$, $2$, and $3$,
since $\kappa$ must always be smaller than $\Omega_\bot$.
When numerical figures are presented, we hereafter take
$\Omega_\bot=\Omega=\Omega_{\rm K}$, where $\Omega_{\rm K}$ is the
relativistic Keplerian angular velocity of rotation.
Then, the radii where resonances occur are $4.0r_{\rm g}$
for ${\tilde n}=1$, $6.0r_{\rm g}$ for ${\tilde n}=2$, and
$12.0r_{\rm g}$ for ${\tilde n}=3$, where   
$r_{\rm g}$ is the Schwarzschild radius defined by
$2GM/c^2$ and $M$ is the mass of the central object.

The spin frequency, $\omega_{\rm s}$, and the frequency of the disk 
oscillations, $\omega$, which make the above resonances possible are
obtained from $\omega_{\rm s}-m_{\rm s}\Omega=\pm\kappa$, and 
$\omega-m\Omega=\pm\kappa$, respectively.
The numerical values of $\omega_{\rm s}$ and the ratio 
$\omega/\omega_{\rm s}$ 
in the case where the mass of a central neutron star, $M$, is $1.4M_\odot$
are shown in tables 1, 2, and 3 for some values of $m$ and $m_{\rm s}$. 
It is noted that the value of $\omega/\omega_{\rm s}$ is independent
of the mass of the central star, while $\omega_{\rm s}$ varies
as $\omega_{\rm s}\propto (M/1.4M_\odot)^{-1}$ if $M$ is changed.

\subsubsection{p-mode oscillations}

There are two kinds of resonances, i.e., horizontal and vertical resonances.
They are separately studied.

\bigskip\noindent
(a) Horizontal resonances

The resonance condition in this case is obtained from the set of 
equations (\ref{2.2}), (\ref{3.2}), and (\ref{3.5}).
Elimination of $\omega$ and $\omega_{\rm s}$ from these equations
shows that the resonance radii are given by 
\begin{equation}
      \kappa={1\over 2}n^{1/2}\Omega_\bot, \quad ({\rm GPH})
\label{4.2}
\end{equation}
where GPH implies that the spin-induced perturbations are 
the g-{\it mode type},
the disk oscillations are {\it p-modes}, and the resonances are {\it horizontal}.
This equation shows that the resonances occur for disk oscillations of
$n=1$, 2, or 3.
The resonance radii are the same as those 
discussed in subsubsection 4.1.1.
Different from the case of subsubsection 4.1.1, 
the disk oscillations must satisfy $\omega-m\Omega=\pm
n^{1/2}\Omega_\bot$ 
and the spin frequencies must satisfy $\omega_{\rm s}-m_{\rm s}\Omega
=\pm \kappa$ at the resonance radii.
The results are also given in tables 1, 2, and 3, as parts of the tables.

\bigskip\noindent
(b) Vertical resonances

Next, we consider the vertical resonances.
The equations to be adopted here are the set of equations
(\ref{2.2}), (\ref{3.2}), and (\ref{3.6}).
Eliminating $\omega$ and $\omega_{\rm s}$ from these equations, we have
\begin{equation}
  \kappa=({\tilde n}^{1/2}-n^{1/2})\Omega_\bot \quad {\rm or} \quad
  \kappa=(n^{1/2}-{\tilde n}^{1/2})\Omega_\bot, \quad ({\rm GPV})
\label{4.3}
\end{equation}
depending on which of ${\tilde n}$ and $n$ is larger.
Here, GPV means that the spin-induced perturbations are the g-{\it mode type}, 
the oscillations are {\it p-mode}, and the resonances are {\it vertical}.
This kind of resonances occurs at many radii.
We, however, restrict here our attention only to $n$, ${\tilde n}=1$, 2, 
and 3, since the cases of large $n$ and ${\tilde n}$ are  
observationally uninteresting.
Then, the radii given by equation (\ref{4.3}) are
$\kappa=(\sqrt{3}-\sqrt{2})\Omega_\bot$, $(\sqrt{3}-1)\Omega_\bot$,
and $(\sqrt{2}-1)\Omega_\bot$.
The spin frequencies, $\omega_{\rm s}$, i.e., $\omega_{\rm s}-m_{\rm s}\Omega
=\pm \kappa$, and the frequencies, $\omega$, of the disk oscillations, 
i.e., $\omega-m\Omega=\pm n^{1/2}\Omega_\bot$, in these cases are 
shown in tables 4, 5, and 6.

\subsection{p-Mode-Type Spin-Induced Perturbations}

In this subsection, we consider the case where the perturbations induced
on disks by the spin of the central star are p-mode-type oscillations,
described by equations (\ref{2.3}).
The two cases concerning disk oscillations, i.e., the cases of g-mode
oscillations and p-mode oscillations, are separately considered below.

\subsubsection{{\rm g}-mode oscillations}

The set of equations to be combined to examine the resonances in this case 
are equations (\ref{2.3}), (\ref{3.1}), and (\ref{3.5}) for 
the horizontal resonances.
When the vertical resonances are studied, however, equation (\ref{3.6}) 
should be used instead of equation (\ref{3.5}).

\bigskip\noindent
(a) Horizontal resonances

First, we consider the horizontal resonances.
Eliminating $\omega_{\rm s}$ and $\omega$ from equations (\ref{2.3}),
(\ref{3.1}), and (\ref{3.5}), we have
\begin{equation}
     \kappa={1\over 2}n_{\rm s}^{1/2}\Omega_\bot, \quad ({\rm PGH})
\label{4.4}
\end{equation}
where PGH implies that the spin-induced perturbations are 
{\it p-mode type},
the disk oscillations are g-{\it modes} and the resonances are 
{\it horizontal}.
The resonances appear only when the spin-induced perturbations have 
$n_{\rm s}=1$, 2, or 3.
The spin frequencies, $\omega_{\rm s}$, satisfying 
$\omega_{\rm s}-m_{\rm s}\Omega=\pm n_{\rm s}^{1/2}\Omega_\bot$
and the frequencies, $\omega$, of the disk oscillations satisfying 
$\omega-m\Omega=\pm\kappa$ are also given in tables 1, 2, and 3.

\bigskip\noindent
(b) Vertical resonances

Next, we consider the vertical resonances.
In this case we solve the set of equations (\ref{2.3}), (\ref{3.1}),
and (\ref{3.6}).
The results show that the resonances occur at radii satisfying 
the relation
\begin{equation}
    \kappa=(n_{\rm s}^{1/2}-{\tilde n}^{1/2})\Omega_\bot \quad {\rm or}\quad
    \kappa=({\tilde n}^{1/2}-n_{\rm s}^{1/2})\Omega_\bot,
    \quad ({\rm PGV})
\label{4.5}
\end{equation}
depending on the magnitude of $n_{\rm s}$ and ${\tilde n}$.
The radii of these resonances are the same as those in the case of GPV,
but the values of $\omega$ and $\omega_{\rm s}$ required for these resonances
are different from those in the case of GPV. 
The results are given in tables 4, 5, and 6.

\subsubsection{p-mode oscillations}

As in the previous cases, there are two possibilities concerning the
types of resonances.
One is a horizontal resonance, and the other is a vertical resonance.
First, we consider the horizontal resonances.
The equations that we should use are equations (\ref{2.3}),
(\ref{3.2}), and (\ref{3.5}).
From these equations, we have
\begin{equation}
     \kappa=(n^{1/2}-n_{\rm s}^{1/2})\Omega_\bot \quad {\rm or} \quad
     \kappa=(n_{\rm s}^{1/2}-n^{1/2})\Omega_\bot,
    \quad ({\rm PPH})
\label{4.6}
\end{equation}
depending which of $n$ and  $n_{\rm s}$ is larger.
The resonances occur at the same radii as in the cases of GPV and 
PGV, but the frequencies of the resonant oscillations are different.
The results are also given in tables 4, 5, and 6.

The final is the case where the resonances are vertical.
The set of equations to be used are equations (\ref{2.3}),
(\ref{3.2}), and (\ref{3.6}).
These equations give
\begin{equation}
     {\tilde n}=n^{1/2}+n_{\rm s}^{1/2}, \quad {\rm or}\quad
       n^{1/2}-n_{\rm s}^{1/2}, \quad {\rm or} \quad
       n_{\rm s}^{1/2}-n^{1/2}.
\label{4.7}
\end{equation}
These relations cannot be satisfied as long as $n$, $n_{\rm s}$, and 
${\tilde n}$ are positive integers.
That is, there is no resonance in this case.

\section{Discussion}

Tables 1--6 give the ratio $\omega/\omega_{\rm s}$ in cases where
the disk oscillations have resonant interactions with the disks through
non-linear coupling with spin-induced perturbations.
The spin-frequency, $\omega_{\rm s}$, in these tables are for 
the mass of the central source being $1.4M_\odot$.
The ratio $\omega/\omega_{\rm s}$ is, however, independent of the mass. 
Resonant interactions occur only for cases where the central 
star has some particular spin frequencies, as shown in tables,
and for combinations of i) the modes of the spin-induced
perturbations [g-mode-type(G) or p-mode-type(P) motion], 
ii) the types of disk oscillations
[g-mode(G) or p-mode(P)], and iii) the types of resonant interactions
[horizontal(H) or vertical(V)].
The tables are restricted only to cases where $m$, $m_{\rm s}$,
$n$, and $n_{\rm s}$, and ${\tilde n}$ are
smaller than 3, except for some particular cases, since the oscillations 
with a large number of nodes in the azimuthal and  
vertical directions would be observationally uninteresting.

Recent observations of kHz QPOs in X-ray binaries show that in neutron-star 
X-ray binaries where the spin frequency is known, 
the frequency difference of a pair kHz QPOs is close to,
or half of, the spin frequency of the central stars, i.e., 
$\Delta\omega\sim\omega_{\rm s}$ or $\Delta\omega\sim\omega_{\rm s}/2$
[van der Klis (2000) for review, and see also Wijnands et al. (2003)].
Their appearance may not be by chance, and
should be explained.
Tables 1--6 show that in our present model there are many cases where
$\Delta\omega=\omega_{\rm s}$, $\Delta\omega=\omega_{\rm s}/2$,
or $\Delta\omega=\omega_{\rm s}/3$ holds exactly.
Such cases are summarized in tables 7--12.
The set of disk oscillations with $\Delta \omega=\omega_{\rm s}$ 
is given in the tables by figures enclosed by parentheses.
The set of disk oscillations with $\Delta\omega=\omega_{\rm s}/2$
is shown by figures enclosed by double brackets.
The set of disk oscillations whose frequency difference is close to
$\Delta\omega=\omega_{\rm s}/2$ (not equal to 
$\Delta\omega=\omega_{\rm s}/2$) is shown by attaching the
superscript * in addition to double brackets.
The cases where $\Delta\omega=\omega_{\rm s}/3$ holds    
are shown in table 7 by triple brackets.

Tables 7--12 show that $\Delta\omega=\omega_{\rm s}$ occurs in 
slowly rotating objects as well as in rapidly rotating ones.
Unlike this, the resonances of $\Delta\omega=\omega_{\rm s}/2$
occur only when the spin of the central object is rather high,
i.e., $\omega_{\rm s}\geq$ 1 kHz (see tables 7, 11, and 12), when 
$M\sim 1.4M_\odot$.
(In the case of $M=1.4M_\odot$, the Kepler frequency at $r=3r_{\rm g}$
is $1.57$ kHz.)
Even if the rotation is not very rapid, 
however, our resonance model has cases of 
$\Delta\omega \cong \omega_{\rm s}/2$ (see table 9).

Let us now discuss individual objects.
Wijnands et al. (2003) recently reported the detection of kHz QPOs
from a millisecond pulsar, SAX J1808.4$-$3658.
They are a pair whose frequencies are close to 500 Hz and 700 Hz.
The spin frequency, on the other hand, is inferred to be 400 Hz.
This object is a rather convincing example of 
$\Delta\omega\sim\omega_{\rm s}/2$,
and the ratio of $\omega/\omega_{\rm s}$ is 1.25 and 1.75.
In our resonance mode, a case where the frequencies of possible two 
oscillations are exactly
1.25 and 1.75 (in units of $\omega_{\rm s}$) exists when 
$\omega_{\rm s}=2035$ Hz (for $M=1.4M_\odot$), as shown in tables 1 and 7.
If we regard this set of pair oscillations as those
observed in SAX J1808.4$-$3658, the difference between 2035 Hz and the 
observed frequency, 400 Hz, must be regarded as mass difference.
The scaling concerning mass gives that 
the mass of the central source of SAX J1808.4$-$3658 is 
$1.4\times(2035/400)M_\odot$, which is $7.1 M_\odot$.
Such a large mass cannot be a neutron star, suggesting that 
this identification is inappropriate. 
In our model, however, there is a case where $\omega/\omega_{\rm s}$ 
is close to 1.25 and 1.75, although they are not exactly equal 
to 1.25 and 1.75.
This is a case of GPH resonances at $r=12.0r_{\rm g}$ with $m_{\rm s}=3$ 
and $n=3$.
That is, table 3 (see also table 9) shows that when the mass of 
the central source is $1.4M_\odot$
and the spin frequency is $\omega_{\rm s}=418$ Hz, there are disk
oscillations whose frequencies are described by 1.28$(m=1)$ and
1.75$(m=2)$ in units of $\omega_{\rm s}$.
They are close to the observed set of ratios, i.e., 1.25 and 1.75.
The difference between 1.28 and 1.25 is not serious, since in our model
the frequencies of disk oscillations are not robust, as discussed later.
Since the observed spin frequency of SAX J1808.4$-$3658 is 400 Hz,
a comparison of 418 Hz in table 3 with the observed 400 Hz leads to the fact 
that the mass of the object is 1.46 $M_\odot[=1.4M_{\odot}\times (418/400)]$.
Wijnands et al. (2003) report that in SAX J1808.4$-$3658 a  
third QPO near 410 Hz was detected on four occasions.
This frequency is 1.03 in units of $\omega_{\rm s}$.
The presence of such an oscillation is not suprising in our model,
since $\omega/\omega_{\rm s}=1.06$ is present as an
oscillation of $m=4$ (see table 3).

Among the five objects listed by van der Klis (2000) as the objects where 
both a pair kHz QPOs and spin frequency are observed, 
KS 1731$-$260 seems to belong to the 
same category as SAX J1808.4$-$3658.
In KS 1731$-$260, the observed spin frequency is 524 Hz and three kHz QPOs 
(900 Hz, 1160 Hz, and 1205 Hz) have been observed.
They are 1.72, 2.21, and 2.30 in units of $\omega_{\rm s}$(=524 Hz).
A set of oscillations close to them are really found in GPH resonances
($m_{\rm s}=3$, $n=3$) 
at $12.0r_{\rm g}$, i.e., 1.75 and 2.22, and in GGV resonances
($m_{\rm s}=3$, ${\tilde n}=3$) at $r=12.0r_{\rm g}$, i.e.,
2.28 (see table 3 and also table 9).
This is realized when the spin frequency is 418 Hz (when $M=1.4M_\odot$).
If we identify these oscillations with the observed ones, the mass of the 
central source of KS 1731$-$260 is $1.4M_\odot\times (418/524)=1.1 M_\odot$.

Let us discuss other sources listed in table 4 of van der Klis (2000).
First we consider 4U 1702$-$43.
Among the various frequencies observed in this source, we pick up 655 Hz
and 1000 Hz.
Since the observed spin frequency is 330 Hz, we have $\omega/\omega_{\rm s}=
1.98$ and 3.03, which are close to 2:3.
In our model, a pair of disk oscillations of 2:3 occurs at $r=4.0r_{\rm g}$ 
(GPH and GGV resonances) when
the spin frequency is 509 Hz (for $M=1.4M_\odot$), 
as shown in tables 1 and 7.
If we identify these oscillations with the observed ones, the mass 
of the central source becomes $2.1M_\odot$ by the scaling of mass, 
i.e., $M=1.4M_\odot\times (509/330)=2.1 M_\odot$.
This mass might be somewhat larger than a possible upper limit of the 
netron-star mass.
There is, however, another possibility that the pair corresponds to 
2.00($m=2$) and 3.08 ($m=4$) of GPH resonances ($m_{\rm s}=1$
and ${\tilde n}=3$) at $12.0r_{\rm g}$ (see table 3). 
The spin frequency in this case is 365 Hz if the central mass is $1.4M_\odot$.
If we adopt this identification, the mass of the central source is
$1.4M_\odot\times (365/330)=1.55 M_\odot$.
This identification is supported if we consider that when
the spin frequency is 365 Hz ($M=1.4M_\odot$), there are  
GGV resonances ($m_{\rm s}=1$ and ${\tilde n}=3$) and we have 
oscillations of 2.07($m=3$) and 2.60($m=4$) (see table 3).
The observations, on the other hand, show the presence of
oscillations of 2.12 (700 Hz) and 2.73 (902 Hz).
They are close to 2.07 and 2.60, respectively. 

In 4U 1728$-$34, many kHz oscillations have been observed.
We take here the oscillations of 510 Hz and 875 Hz as typical ones.
Then, we have $\omega/\omega_{\rm s}=1.40$ and 2.40, since the
spin frequency, $\omega_{\rm s}$, is known to be 364 Hz.
Table 2 shows that in our resonance model, oscillations of 
$\omega/\omega_{\rm s}=1.41 (m=1)$ and $2.41(m=0)$ are present, respectively,
in GPH (with $m_{\rm s}=1$ and $n=2$) and GGV (with $m_{\rm s}=1$ and
${\tilde n}=2$) resonances at $r=6.0r_{\rm g}$.
The required spin frequency is 162 Hz when the mass is $1.4M_\odot$.
If we adopt this identification, the mass of this object is
$1.4M_\odot \times(162/364)=0.62M_\odot$.

In the case of 4U 1636$-$53 we adopt 900 Hz and 1190 Hz as typical
oscillations.
The spin frequency, $\omega_{\rm s}$, is 582 Hz.
Hence, $\omega/\omega_{\rm s}=1.55$ and 2.04, and
$\Delta\omega\sim\omega_{\rm s}/2$.
In our model, GGV resonances (with $m_{\rm s}=1$ and ${\tilde n}=3$) 
at $r=12.0r_{\rm g}$ have oscillations
of 1.54($m=2$) and 2.07($m=3$) in units of $\omega_{\rm s}$, 
as shown in table 3.
The spin frequency for these oscillations is 365 Hz when 
$M=1.4M_\odot$.
If we adopt this identification, the mass of 4U 1536$-$53 is 
$1.4M_\odot\times(365/582)=0.87M_\odot$.

Finally, we consider Aq1X-1, which has oscillations of 
670 Hz, 930 Hz, and 1040 Hz.
Since the spin frequency is 549 Hz, the frequencies of these
oscillations are 1.22, 1.69, and 1.89 in units of the spin frequency.
In our model, we have $1.23(m=3)$, $1.66(m=2)$, and $1.81(m=4)$
for PGH resonances (with $m_{\rm s}=0$ and $n_{\rm s}=3$) at 
$r=12.0r_{\rm g}$ (see table 3).
The required spin frequency for these resonances is 339 Hz 
when $M=1.4M_\odot$.
If we regard the observed QPOs as those mentioned above,
the mass of the central object is
$1.4M_\odot\times(339/549)=0.86M_\odot$.
All of the results mentioned above for six objects are summarized in
table 13.

It is encouraging that only a few types of resonances appear in table 13.
In principle, there are six types of resonances, i.e.,
GGV, GPH, GPV, PGH, PGV, PPH (GGH and PPV are absent).
Each one is further divided into subclasses by different combinations of
$m_{\rm s}$, $n_{\rm s}$, $n$, etc.
In spite of these varieties of resonances, all QPOs in two sources 
(J1808.4$-$3658 and KS 1731$-$260) and some of QPOs in three sources
(4U 1702$-$43, 4U 1728$-$34, 4U 1636$-$53) are accomodated with the types of
GPH and GGV.
This suggests that this accomodation is not by chance,
but these resonances really work in many sources.

Here, we discuss some characteristics of the present resonant model.
First, we should emphasize that for our present resonances to work,
the central star must have some particular spin frequencies, as
shown in the tables.
This does not mean that all sources with kHz QPOs must have particular
spin frequencies.
Our standpoint is that kHz QPOs come from resonant interactions
of disk oscillations with deformed disks.
As possible deformations of disks there are some cases.
The most interesting one in relation to kHz QPOs is a warp, and we
proposed models that kHz QPOs in neutron-star and high-frequency QPOs
in black-hole X-ray binaries are due to resonant interactions of 
disk oscillations with warped disks (Kato 2005a,b).
The present resonances in spin-induced deformed-disks are supplementary 
in the sense that they work only when spin has some particular values.
 
In the present resonance model, as in the warp model, the appearance
of oscillations in a pair is not essential.
This is favorable for the fact that observed QPOs are not always a pair,
but occasionally only one QPO is observed.
In our model, however, the cases where three or more oscillations
simultaneously appear are not excluded.
For example, a set of oscillations of different azimuthal
wavenumber $m$, with the same other parameters are possible.
The QPOs in J1808.4$-$3658 and AqlX-1 can be regarded as such a set of
oscillations (see table 13).
One of tests whether our present model really works is to examine
whether there are other sources where more than three QPOs are observed, 
and they can be regarded as
a set of oscillations with different $m$, although other parameters
are the same.

Comparison of QPO frequencies derived by our present model to those of 
observations showed that in some cases their agreements are not good enough.
It is important to note here that these differences are natural and
rather favorable for the present model, since in the present model
the temperature in disks is assumed to be isothermal in the
vertical direction.
In real disks, this will not be the case.
The temperature may decrease in the vertical direction.
Also, it may change with time.
Then, the frequencies of p-mode oscillations with $n\not= 1$ as well 
as the conditions of vertical resonances of ${\tilde n}\not= 1$ are
modified (Kato 2005b).
Considerations of these modifications may more favarably explain the observed
QPO frequencies and their time variations. 
Further examinations in this direction are worthwhile in the future.


\newpage

\begin{longtable}{rrrrrrrrr}

  \caption{Frequencies of resonant oscillations 
    at $\kappa=(1/2)\Omega$ $(r=4.0r_{\rm g})$
    in units of $\omega_{\rm s}$.}
\label{table 1}
\endfirsthead

\hline\hline

  GGV &

\multicolumn{4}{c} {$\omega=m\Omega+\kappa$} &

\multicolumn{4}{c} {$\omega=m\Omega-\kappa$} \\

  $\omega_{\rm s}(m_{\rm s}, {\tilde n})$  & 

  $m=0$&$m=1$&$m=2$& $m=3$&$m=0$&$m=1$ & $m=2$& $m=3$\\

\hline

  $\mp$509Hz(0,1) & $\mp$1.00 & $\mp$3.00 & $\mp$5.00 & $\mp 7.00$ & 

              $\pm$1.00 & $\mp$1.00 & $\mp$3.00 & $\mp$5.00 \\

  509Hz(1,1) & 1.00 & 3.00 & 5.00 & 7.00 & 

               $-1.00$ & 1.00 & 3.00 & 5.00 \\

  1526Hz(2 or 1,1) & 0.33 & 1.00 & 1.67 & 2.33 & 

               $-0.33$ & 0.33 & 1.00 & 1.67 \\

\hline

 GPH &

\multicolumn{4}{c} {$\omega=m\Omega+n^{1/2}\Omega_\bot$} & 

\multicolumn{4}{c} {$\omega=m\Omega-n^{1/2}\Omega_\bot$}   \\

  $\omega_{\rm s}(m_{\rm s}, n)$  & 

  $m=0$&$m=1$&$m=2$& $m=3$& $m=0$ &$m=1$&$m=2$& $m=3$ \\

\hline

  $\mp$509Hz(0,1) & $\mp$2.00 & $\mp$4.00 & $\mp$6.00 & $\mp 8.00$ & 
  $\pm$2.00 & 0.00 & $\mp$2.00 & $\mp4.00$ \\

  509Hz(1,1) & 2.00 & 4.00 & 6.00 & 8.00 & 
              $-2.00$ & 0.00 & 2.00 & 4.00 \\

  1526Hz(1,1) & 0.67 & 1.33 & 2.00 & 2.67 & 
              $-0.67$ & 0.00 & 0.67 & 1.33  \\

\hline

 PGH &
\multicolumn{4}{c} {$\omega=m\Omega+\kappa$} &  

\multicolumn{4}{c} {$\omega=m\Omega-\kappa$}  \\

  $\omega_{\rm s}(m_{\rm s}, n_{\rm s})$  & 

  $m=0$&$m=1$&$m=2$& $m=3$& $m=0$ &$m=1$&$m=2$& $m=3$ \\

\hline

  0Hz(1,1)  & ... & ... & ... & ... & ... & ... & ... & ... \\

  $\mp$1018Hz(0,1) & $\mp$0.50 & $\mp$1.50 & $\mp$2.50 & $\mp$ 3.50 & 
  $\pm$0.50 & $\mp$0.50 & $\mp$1.50 & $\mp 2.50$ \\

  1018Hz(2,1) & 0.50 & 1.50 & 2.50 & 3.50& 
          $-0.50$ & 0.50 & 1.50 & 2.50 \\

  2035Hz(1,1) & 0.25 & 0.75 & 1.25 & 1.75 & 
          $-0.25$ & 0.25 & 0.75 & 1.25 \\

\hline
\end{longtable}

\begin{longtable}{rrrrrrrrr}
  \caption{Frequencies of resonant oscillations 
   at $\kappa=(\sqrt{2}/2)\Omega$ $(r=6.0r_{\rm g})$
   in units of $\omega_{\rm s}$.}
\label{table 2}
\endfirsthead

\hline\hline

  GGV &
\multicolumn{4}{c} {$\omega=m\Omega+\kappa$} &

\multicolumn{4}{c} {$\omega=m\Omega-\kappa$}   \\

  $\omega_{\rm s}(m_{\rm s}, {\tilde n})$  & 
  $m=0$&$m=1

$&$m=2$&$m=3$& $m=0$ &$m=1$&$m=2$& $m=3$ \\
\hline
  162Hz(1,2) & 2.41 & 5.83 & 9.24 & 12.65& $-2.41$ & 1.00 

              & 4.42 & 7.77  \\

 $\mp$392Hz(0,2) & $\mp$1.00 & $\mp$2.41 & $\mp$3.83 & $\mp 5.24$ 

      & $\pm$1.00 & $\mp 0.41$ & $\mp 1.83$  & $\mp 3.24$  \\

 716Hz(2,2) & 0.55 & 1.32 & 2.09 & 2.86 & $-0.55$ & 0.23 & 1.00 

      & 2.45 \\

 946Hz(1,2) & 0.41 & 1.00 & 1.59 & 2.17 & $-0.41$ & 0.17 & 0.76 

      & 1.34  \\

 1270Hz(3,2) & 0.31 & 0.74 & 1.18 & 1.62 & $-0.31$ & 0.13 & 0.56 

      & 1.00  \\

 1500Hz(2,2) & 0.26 & 0.63 & 1.00 & 1.37 & $-0.26$ & 0.11 & 0.48 

      & 0.85  \\

 2054Hz(3,2) & 0.19 & 0.46 & 0.73 & 1.00 & $-0.19$ & 0.08 & 0.35 

      & 0.62 \\ 

\hline

 GPH &

\multicolumn{4}{c} {$\omega=m\Omega+n^{1/2}\Omega_\bot$} &

\multicolumn{4}{c} {$\omega=m\Omega-n^{1/2}\Omega_\bot$}  \\

  $\omega_{\rm s}(m_{\rm s}, n)$  & 
  $m=0$&$m=1$&$m=2$&$m=3$& $m=0$&$m=1$&$m=2$& $m=3$ \\
\hline
  162Hz(1,2) & 4.83 & 8.24 & 11.65& 15.07 & $-4.83$ & $-1.41$ & 2.00 

             & 5.41 \\
  $\mp$392Hz(0,2)  & $\mp$2.00 & $\mp$3.41 & $\mp$4.83 & $\mp 6.24$ 

             & $\pm$2.00 & $\pm$0.59 & $\mp0.83$ & $\mp2.24$ \\

  716Hz(2,2) & 1.09 & 2.30 & 2.64 & 3.41 & $-1.09$ & $-0.32$& 0.45 & 1.23 \\

  946Hz(1,2) & 0.83 & 1.41 & 2.00 & 2.59 & $-0.83$ & $-0.24$& 0.34 & 0.93 \\
 1270Hz(3,2) & 0.62 & 1.18 & 1.49 & 1.93 & $-0.62$ & $-0.18$ & 0.26 

             & 0.69  \\

 1500Hz(2,2) & 0.52 & 0.89 & 1.26 & 1.63 & $-0.52$ & $-0.15$ & 0.22 

             & 0.59  \\

 2054Hz(3,2) & 0.38 & 0.65 & 0.92 & 1.19 & $-0.38$ & $-0.11$ & 0.15

             & 0.43 \\

\hline

 PGH &

\multicolumn{4}{c} {$\omega=m\Omega+\kappa$} &

\multicolumn{4}{c} {$\omega=m\Omega-\kappa$}   \\

  $\omega_{\rm s}(m_{\rm s}, n_{\rm s})$  & 

  $m=0$&$m=1$&$m=2$&$m=3$& $m=0$ &$m=1$&$m=2$ & $m=3$\\

\hline

  $-229$Hz(1,2) & $-1.71$ & $-4.12$ & $-6.54$& $-8.95$ & $1.71$ & $-0.71$

     & $-3.12$  & $-5.54$ \\

  325Hz(2,2) & 1.21 & 2.90 & 4.62 & 6.33 & $-1.21$ & 0.50 & 2.21 & 3.91 \\

  $\mp$554Hz(0,2) & $\mp$0.71 & $\mp$1.71 & $\mp$2.71& $\mp3.71$

       & $\pm$0.71 & $\mp$0.29& $\mp$1.29 & $\mp$2.29 \\

  879Hz(3,2) & 0.45 & 1.08 & 1.71& 2.34 & $-0.45$ & 0.19& 0.82 & 1.45 \\

 1337Hz(1,2) & 0.29 & 0.71 & 1.26& 1.54 & $-0.29$ & 0.12& 0.54 & 0.95 \\

\hline

\end{longtable}

\begin{longtable}{rrrrrrrrrrr}
  \caption{Frequencies of resonant oscillations
   at $\kappa=(\sqrt{3}/2)\Omega$ $(r=12.0r_{\rm g})$
   in units of $\omega_{\rm s}$.}
\label{table 3}
\endfirsthead
\hline\hline
  GGV &

 \multicolumn{5}{c} {$\omega=m\Omega+\kappa$} & 
 \multicolumn{5}{c} {$\omega=m\Omega-\kappa$}  \\

  $\omega_{\rm s}(m_{\rm s},{\tilde n})$  
  & $m=0$&$m=1$&$m=2$&$m=3$&$m=4$& $m=0$ &$m=1$&$m=2$&$m=3$&$m=4$
   \\
\hline
  26.2Hz(1,3) & 6.47 & 13.93 & ... & ... & ... & $-6.47$& 1.00 

         & 8.47 & 15.92 & ...\\

  $\mp$170Hz(0,3) & $\mp$1.00 & $\mp$2.15 & $\mp$3.31 &$\mp$4.46

         & $\mp$5.62 & $\pm$1.00 

         & $\mp$0.15 & $\mp$1.31 & $\mp$2.46 & $\mp$3.62\\

  222Hz(2,3) & 0.76 & 1.65 & 2.53 & 3.41& 4.29 & $-0.76$ 

         & 0.12 & 1.00 & 1.88 & 2.76 \\

  365Hz(1,3) & 0.46 & 1.00 & 1.54 & 2.07& $2.61$ & $-0.46$ 

         & 0.07 & 0.61 & 1.14 & 1.68 \\

  418Hz(3,3) & 0.41 & 0.87 & 1.34 & 1.81& 2.28 & $-0.41$ 

         & 0.06 & 0.53 & 1.00 & 1.47 \\

  561Hz(2,3) & 0.30 & 0.65 & 1.00 & 1.35& 1.70 & $-0.30$ 

         & 0.05 & 0.43 & 0.75 & 1.07 \\

  757Hz(3,3) & 0.22 & 0.48 & 0.74 & 1.00 & 1.26 & $-0.22$

         & 0.03 & 0.29 & 0.55 & 0.81 \\

\hline 
  GPH &

\multicolumn{5}{c} {$\omega=m\Omega+n^{1/2}\Omega_\bot$} &
\multicolumn{5}{c} {$\omega=m\Omega-n^{1/2}\Omega_\bot$} \\

  $\omega_{\rm s}(m_{\rm s},n)$  
  & $m=0$&$m=1$&$m=2$&$m=3$& $m=4$ & $m=0$&$m=1$&$m=2$&$m=3$&$m=4$\\
\hline
  26.2Hz(1,3) & 12.9 & 20.4 & ... & ... & ... & $-12.9$& $-5.46$ 

              & 1.15 & 9.46 & ... \\

  $\mp$170Hz(0,3) & $\mp$2.00 & $\mp$3.15 & $\mp$4.31 & $\mp 5.46$ 

     & $\mp$6.62 & $\pm$2.00 & $\pm$0.85 & $\mp$0.31 & $\mp 1.46$ 

     & $\mp$2.62 \\
  222Hz(2,3) & 1.53 & 2.41 & 3.29 & 4.17& 5.05 & $-1.53$ & $-0.65$ 

             & 0.24 & 1.12 & 2.00 \\
  365Hz(1,3) & 0.93 & 1.46 & 2.00 & 2.54  & 3.07 & $-0.93$ & $-0.39$ 

             & 0.14 & 0.68 & 1.22 \\
  418Hz(3,3) & 0.81 & 1.28 & 1.75 & 2.22& 2.69 & $-0.81$ & $-0.34$ 

             & 0.13 & 0.59 & 1.06 \\
  561Hz(2,3) & 0.60 & 0.95 & 1.30 &1.65 & 2.00 & $-0.60$ & $-0.26$ 

             & 0.09 & 0.44 & 0.79 \\

  757Hz(3,3) & 0.45 & 0.71 & 0.97 & 1.22 & 1.48 & $-0.45$ & $-0.19$ 

             & 0.07 & 0.33 & 0.59 \\
\hline       
  PGH &

\multicolumn{5}{c} {$\omega=m\Omega+\kappa$} & 
\multicolumn{5}{c} {$\omega=m\Omega-\kappa$}  \\

   $\omega_{\rm s}(m_{\rm s},n_{\rm s})$  
  & $m=0$& $m=1$& $m=2$& $m=3$& $m=4$ &$m=0$& $m=1$& $m=2$& $m=3$ &$m=4$ \\
\hline

  52.5Hz(2,3)& 3.23& 6.96 & 10.69 &14.43& ... & $-3.23$ & 0.50 & 4.23 & 7.96 & ... \\

  $-143$Hz(1,3)  & $-1.18$ & $-2.55$ & $-3.92$ & $-5.28$ 

                 & -6.65 & 1.18 & $-0.18$ 

                 & $-1.55$ & $-2.92$ & -4.28 \\
  248Hz(3,3) &0.68 & 1.47 & 2.26 & 3.05 & 3.84 & $-0.68$ & 0.11 & 0.89 

                 & 1.68 & 2.47 \\

  $\mp 339$Hz(0,3) & $\mp 0.50$ & $\mp 1.08$ & $\mp 1.66$ & $\mp 2.23$ 

             & $\mp$2.81 

             & $\pm 0.50$ & $\mp 0.08$ & $\mp 0.66$ & $\mp 1.23$&$\mp 1.81$ \\
  535Hz(1,3) & 0.32 & 0.68 & 1.05 & 1.42 & 1.78 &  $-0.32$ & 0.05 & 0.42 & 0.78 

             & 1.15 \\

  731Hz(2,3) & 0.23 & 0.50 & 0.77 & 1.04 & 1.30 & $-0.23$ & 0.04 & 0.30 & 0.57 

             & 0.84 \\

  927Hz(3,3) & 0.18 & 0.39 & 0.61 & 0.82 & 1.03 & $-0.18$ & 0.03 & 0.24 & 0.45 

             & 0.66  \\
\hline
\end{longtable}

\begin{longtable}{rrrrrrrrr}

\caption{Frequencies of resonant oscillations 
 at $\kappa=(\sqrt{3}-\sqrt{2})\Omega$ $(r=3.34r_{\rm g})$
 in units of $\omega_{\rm s}$.}
\label{table 4}

\endfirsthead

\hline\hline

  GPV &

\multicolumn{4}{c} {$\omega=m\Omega+n^{1/2}\Omega_\bot$} & 

\multicolumn{4}{c} {$\omega=m\Omega-n^{1/2}\Omega_\bot$}   \\

  $\omega_{\rm s}(m_{\rm s}, n, {\tilde n})$  & 

  $m=0$&$m=1$&$m=2$&$m=3$&$m=0$&$m=1$ & $m=2$& $m=3$ \\

\hline

  $\mp$424Hz(0,2,3) & $\mp$4.45 & $\mp$7.60 & $\mp$10.74 

      & $\mp 13.89$  & $\pm$4.45 & $\pm$1.30 & $\mp$1.84 & $\mp 4.99$ \\

  909Hz(1,2,3) & 2.07 & 3.54 & 5.00 & 6.47 & $-2.07$ & $-0.61$ & 0.86 & 2.32 \\

 1757Hz(1,2,3) & 1.07 & 1.83 & 2.59 & 3.35 & $-1.07$ & $-0.31$ & 0.45 & 1.20 \\

 2242Hz(2,2,3) & 0.84 & 1.44 & 2.03 & 2.62 & $-0.84$ & $-0.25$ & 0.35 & 0.94 \\

 $\mp$424Hz(0,3,2) & $\mp$5.45 & $\mp$8.60 & $\mp$11.74 & $\mp 14.89$ 

       & $\pm 5.45$ & $\pm$2.30 & $\mp$0.84 & $\mp 3.99$ \\

  909Hz(1,3,2)  & 2.54 & 4.00 & 5.47 & 6.94 & 2.54 & 1.07 & 0.39 & 1.86 \\

  1757Hz(1,3,2) & 1.31 & 2.07 & 2.83 & 4.32 & 1.31 & 0.56 & 0.20 & 0.96 \\

  2242Hz(2,3,2) & 1.03 & 1.62 & 2.22 & 2.81 & 1.03 & 0.44 & 0.16 & 0.75 \\

\hline

 PGV  &

\multicolumn{4}{c} {$\omega=m\Omega+\kappa$} & 

\multicolumn{4}{c} {$\omega=m\Omega-\kappa$}  \\

  $\omega_{\rm s}(m_{\rm s}, n_{\rm s}, {\tilde n})$  & 

  $m=0$&$m=1$&$m=2$&$m=3$&$m=0$&$m=1$ & $m=2$& $m=3$ \\

\hline

  357Hz(2,3,2) & 1.19 & 4.92 & 8.65 & 12.38 & $-1.19$ & 2.55 & 6.28 & 10.01 \\

  $-976$Hz(1,3,2) & $-0.43$ & $-1.80$ & $-3.17$ & $-4.53$ & 0.43 & $-0.93$ & 

                  $-2.30$ & $-3.66$ \\

  1690Hz(3,3,2)& 0.25 & 1.04 & 1.83 & 2.62 & $-0.25$ & 0.54 & 1.33 & 2.12 \\

  $\mp 2308$Hz(0,3,2) & $\mp 0.18$ & $\mp 0.76$ & $\mp 1.34$ & $\mp 1.92$  

         & $\pm 0.18$ & $\mp 0.39$ & $\mp 0.97$ & $\mp 1.55$ \\

  $-552$Hz(1,2,3) & $-0.77$ & $-3.18$ & $-5.60$ & $-8.01$ & 0.77 & $-1.65$ 

         & $-4.06$ & $-6.48$ \\

  781Hz(2,2,3) & 0.54 & 2.25 & 3.96 & 5.66 & $-0.54$ & 1.17 & 2.87 & 4.58 \\

  $\mp 1985$Hz(0,2,3) & $\mp 0.22$ & $\mp 0.93$ & $\mp 1.64$ & $\mp 2.35$ & 

           $\pm 0.22$ & $\mp 0.48$ & $\mp 1.19$ & $\mp 1.90$ \\

  2113Hz(3,2,3)& 0.20 & 0.83 & 1.46 & 2.09 & $-0.20$ & 0.43 & 1.06 & 1.69 \\

\hline

 PPH &
\multicolumn{4}{c} {$\omega=m\Omega+n^{1/2}\Omega_\bot$} & 

\multicolumn{4}{c} {$\omega=m\Omega-n^{1/2}\Omega_\bot$}  \\
  $\omega_{\rm s}(m_{\rm s}, n_{\rm s}, n)$  & 
  $m=0$&$m=1$&$m=2$&$m=3$&$m=0$&$m=1$ & $m=2$& $m=3$ \\
\hline
 $-552$Hz(1,2,3) & $-4.18$ & $-6.60$ & $-9.01$ & $-11.43$& 4.18 & 1.77 

                     & $-0.65$ & $-3.06$ \\

 781Hz(1,2,3) & 2.96 & 4.66 & 6.37 & 8.08 & $-2.96$ & $-1.25$ & 0.46 & 2.16 \\

 $\mp$1985Hz(0,2,3)  & $\mp1.22$ & $\mp1.93$ & $\mp2.64$ & $\mp3.35$  

                     & $\pm1.22$ & $\pm0.52$ & $\mp0.19$ & $\mp0.90$ \\

 2113Hz(3,2,3) & 1.09 & 1.72 & 2.35 & 2.98 & $-1.09$ & $-0.46$ & 0.17 & 0.80 \\

 357Hz(2,3,2) & 5.27 & 9.01 & 12.7 & 16.47 & $-5.27$ & $-1.55$  & 2.19 & 5.92 \\

 $-976$Hz(1,3,2) & $-1.93$ & $-3.30$ & $-4.66$ & $-6.03$ & 1.93 & 0.57 

                  & $-0.80$ & $-2.17$ \\
 $\mp 2308$Hz(0,3,2) & $\mp0.82$ & $\mp1.39$ & $\mp1.97$ & $\mp 2.55$ 

                     & $\pm0.82$ & $\pm0.24$ & $\mp0.34$ & $\mp 0.92$ \\
\hline 
\end{longtable}

\begin{longtable}{rrrrrrrrr}

  \caption{Frequencies of resonant oscillations 
    at $\kappa=(\sqrt{2}-1)\Omega$ ($r=3.62r_{\rm g}$)
     in units of $\omega_{\rm s}$.}
\label{table 5}

\endfirsthead

\hline\hline

  GPV &

\multicolumn{4}{c} {$\omega=m\Omega+n^{1/2}\Omega_\bot$} &  

\multicolumn{4}{c} {$\omega=m\Omega-n^{1/2}\Omega_\bot$}   \\

  $\omega_{\rm s}(m_{\rm s}, n, {\tilde n})$  & 

  $m=0$&$m=1$&$m=2$&$m=3$&$m=0$&$m=1$ & $m=2$& $m=3$ \\

\hline

  $\mp$489Hz(0,1,2)& $\mp$2.41 & $\mp$4.83 & $\mp$7.24 & $\mp 9.66$  

      & $\pm$2.41 & 0.00 & $\mp$2.41 & $\mp$4.83 \\

  691Hz(1,1,2)  & 1.71 & 3.41 & 5.12 & 6.83 & $-1.71$ & 0.00 & 1.71 & 3.41 \\

  1669Hz(1,1,2) & 0.71 & 1.41 & 2.12 & 2.83 & $-0.71$ & 0.00 & 0.71 & 1.41 \\

  1871Hz(2,1,2) & 0.63 & 1.26 & 1.89 & 2.52 & $-0.63$ & 0.00 & 0.63 & 1.26 \\

  2849Hz(2,1,2) & 0.41 & 0.83 & 1.24 & 1.66 & $-0.41$ & 0.00 & 0.41 & 0.83 \\

  $\mp$489Hz(0,2,1)& $\mp$3.41 & $\mp$5.83 & $\mp$8.24 & $\mp10.66$  

      & $\pm$3.41 & $\pm$1.00 & $\mp$1.41 & $\mp$3.83 \\

  691Hz(1,2,1) & 2.41 & 4.12 & 5.83 & 7.54 & $-2.41$ & $-0.71$ & 1.00 & 2.71 \\

  $1669$Hz(1,2,1)& $1.00$ & $1.71$ & $2.41$ & $1.41$ 

      & $1.00$ & $ 0.29$ & $ 0.41$ & $1.12$ \\

  1871Hz(2,2,1)& 0.89 & 1.52 & 2.15 & 2.78 & $-0.89$ & $-0.26$ & 0.37 & 1.00\\

\hline

    PGV &

\multicolumn{4}{c} {$\omega=m\Omega+\kappa$} &  

\multicolumn{4}{c} {$\omega=m\Omega-\kappa$}   \\

  $\omega_{\rm s}(m_{\rm s}, n_{\rm s}, {\tilde n})$  & 

  $m=0$&$m=1$&$m=2$&$m=3$&$m=0$&$m=1$ & $m=2$& $m=3$ \\

\hline

  $-489$Hz(1,2,1) & $-1.00$ & $-3.41$ & $-5.83$ & $-8.24$ & 1.00 & $-1.41$ 

                  & $-3.83$ & $-6.24$ \\

   691Hz(2,2,1) & 0.71  & 2.41  & 4.12  & 5.83 & $-0.71$ & 1.00  & 2.71 & 4.41 \\

  $\mp$1669Hz(0,2,1) & $\mp$0.29 & $\mp$1.00 & $\mp$1.71 & $\mp 2.41$  

      & $\pm$0.29 & $\mp$0.41 & $\mp$1.12 & $\mp 1.12$ \\

  1871Hz(3,2,1) & 0.26 & 0.89 & 1.52 & 2.15 & $-0.26$ & 0.33 & 1.00 & 1.63 \\

  2849Hz(1,2,1) & 0.17 & 0.59 & 1.00 & 1.41 & $-0.17$ & 0.24 & 0.66 & 1.07 \\

  0Hz(1,1,2)    & ... & ... & ... & ... & ... & ... & ... & ... \\

  $\mp$1180Hz(0,1,2) & $\mp$0.41 & $\mp$1.44 & $\mp$2.41 & $\mp 3.41$ & $\pm 0.41$

       & $\mp$0.59 & $\mp$1.59  & $\mp 2.59$ \\

  2360Hz(2 or 1,1,2) & 0.21 & 0.71 & 1.21 & 1.74 & $-0.21$ & 0.29 & 0.79 & 1.29 \\


\hline

PPH &

\multicolumn{4}{c} {$\omega=m\Omega+n^{1/2}\Omega_\bot$} &  

\multicolumn{4}{c} {$\omega=m\Omega-n^{1/2}\Omega_\bot$}   \\

  $\omega_{\rm s}(m_{\rm s}, n_{\rm s}, n)$  & 

  $m=0$&$m=1$&$m=2$&$m=3$&$m=0$&$m=1$ & $m=2$& $m=3$ \\

\hline

  $-489$Hz(1,2,1) & $-2.41$ & $-4.83$ & $-7.24$ & $-9.66$ & 2.41 & 0.00 

                  & $-2.41$ & $-4.83$ \\

   691Hz(2,2,1) & 1.71 & 3.41 & 5.12 & 6.83 & $-1.71$ & 0.00 & 1.71 & 3.41 \\

  $\mp$1669Hz(1,2,1) & $\mp0.71$ & $\mp$1.41 & $\mp$2.12 & $\mp 2.83$ & $\pm 0.71$

                & 0.00 & $\mp$ 0.71 & $\mp$1.41 \\

  1871Hz(3,2,1) & 0.63 & 1.26 & 1.89 & 2.52 & $-0.63$ & 0.00 & 0.63 & 1.26 \\

  2849Hz(1,2,1) & 0.41 & 0.83 & 1.24 & 1.66 & $-0.41$ & 0.00 & 0.41 & 0.83 \\

  0Hz(1,1,2)  & ... & ... & ... & ... & ... & ... & ... & ... \\

  $\mp$1180Hz(0,1,2) & $\mp$1.41 & $\mp$2.41 & $\mp$3.41 &

        $\mp 4.41$ & $\pm$1.41 & $\pm$0.41 & $\mp$0.59 & $\mp$1.59  \\

  1180Hz(2,1,2) & 1.41 & 2.41 & 3.41 & 4.41 & $-1.41$ & $-0.41$ & 0.59 & 1.59 \\

  2360HZ(3 or 1,1,2) & 0.71 & 1.21 & 1.71 & 2.21 & $-0.71$ & $-0.21$ & 0.29 

        & 0.79 \\



\hline

\end{longtable}

\begin{longtable}{rrrrrrrrr}

  \caption{Frequencies of resonant oscillations 
    at $\kappa=(\sqrt{3}-1)\Omega$ $(r=6.46r_{\rm g})$
    in units of $\omega_{\rm s}$.}
\label{table 6}

\endfirsthead

\hline\hline

  GPV &

\multicolumn{4}{c} {$\omega=m\Omega+n^{1/2}\Omega_\bot$} &

\multicolumn{4}{c} {$\omega=m\Omega-n^{1/2}\Omega_\bot$}  \\

  $\omega_{\rm s}(m_{\rm s}, n, {\tilde n})$  & 

  $m=0$&$m=1$&$m=2$&$m=3$&$m=0$&$m=1$ & $m=2$& $m=3$ \\

\hline
  133Hz(1,1,3) & 3.73 & 7.46 & 11.19 & 14.93 & $-3.72$ & 0.00 & 3.72 & 7.46 \\

  $\mp$362Hz(0,1,3) & $\mp$1.37 & $\mp$2.73 & $\mp$4.10 & $\mp$5.46
        & $\pm$1.37 & 0.00      & $\mp$1.37 & $\mp$ 2.73  \\ 
    628Hz(2,1,3) & 0.79 & 1.58 & 2.37 & 3.15 & $-0.58$ & 0.00 & 0.79 & 1.58 \\

  857Hz(1,1,3) & 0.42 & 1.00 & 1.58 & 2.15 & $-0.42$ & 0.00 & 0.42 & 1.00 \\

 1123Hz(3,1,3) & 0.32 & 0.76 & 1.20 & 1.65 & $-0.32$ & 0.00 & 0.32 & 0.76 \\

 1352Hz(2,1,3) & 0.27 & 0.63 & 1.00 & 1.37 & $-0.27$ & 0.00 & 0.27 & 0.63 \\

 1847Hz(3,1,3) & 0.20 & 0.46 & 0.73 & 1.00 & $-0.20$ & 0.00 & 0.20 & 0.46 \\

  133Hz(1,3,1) & 6.46 & 10.19&13.93 &17.66 & $-6.46$ & $-2.73$ & 1.00 & 4.73 \\

 $\mp$362Hz(0,3,1) & $\mp$2.37 & $\mp$3.73 & $\mp$5.10 & $\mp$6.46  

        & $\pm$2.37 & $\pm$1.00 & $\mp$0.37 & $\mp$ 1.73  \\

  628Hz(2,3,1) & 1.37 & 2.15 & 2.94 & 4.27 & $-1.36$ & $-0.58$ & 0.21 & 1.00 \\

  857Hz(1,3,1) & 1.00 & 1.58 & 2.15 & 2.73 & $-1.00$ & $-0.42$ & 0.15 & 0.73 \\

 1123Hz(3,3,1) & 0.76 & 1.20 & 1.65 & 2.09 & $-0.76$ & $-0.32$ & 0.12 & 0.56 \\

 1352Hz(2,3,1) & 0.63 & 1.00 & 1.37 & 1.73 & $-0.63$ & $-0.27$ & 0.10 & 0.46 \\

 1847Hz(3,3,1) & 0.46 & 0.73 & 1.00 & 1.27 & $-0.46$ & $-0.20$ & 0.07 & 0.34 \\

\hline

 PGV &

\multicolumn{4}{c} {$\omega=m\Omega+\kappa$} &

\multicolumn{4}{c} {$\omega=m\Omega-\kappa$}   \\

  $\omega_{\rm s}(m_{\rm s}, n_{\rm s}, {\tilde n})$  & 

  $m=0$&$m=1$&$m=2$&$m=3$&$m=0$&$m=1$ & $m=2$& $m=3$ \\

\hline

  0Hz(1,1,3) & ... & ... & ... & ... & ... & ... & ... & ... \\

 $\mp$495Hz(0,1,3) & $\mp$0.73 & $\mp$1.73 & $\mp$2.73 & $\mp$3.73

       & $\pm$0.73 & $\mp$0.27 & $\mp$1.27 & $\mp$2.27 \\ 

    990Hz(1 or 3,1,3) & 0.37 & 0.87 & 1.37 & 1.87 & $-0.37$ & 0.13 & 0.63 & 1.13 \\

 1485Hz(2,1,3) & 0.24 & 0.58 & 0.91 & 1.24 & $-0.24$ & 0.09 & 0.42 & 0.76 \\

 1980Hz(3,1,3) & 0.18 & 0.43 & 0.68 & 0.93 & $-0.18$ & 0.07 & 0.32 & 0.57 \\

 133Hz(2,3,1) & 2.73 & 6.46 & 10.19 & 13.93 & $-2.73$ & 1.00 & 4.73 & 8.46 \\

 $-362$Hz(1,3,1) & $-1.00$ & $-2.37$ & $-3.73$ & $-5.10$ & $1.00$ 

          & $-0.37$ & $-1.73$ & $-3.10$ \\

  628Hz(3,3,1) & 0.58 & 1.37 & 2.15 & 2.94 & $-0.58$ & 0.21 & 0.54 & 1.79 \\

  $\mp$857Hz(0,3,1) & $\mp$0.42 & $\mp$1.00 & $\mp$1.58 &  $\mp 2.15$

         & $\pm$0.42 & $\mp$0.15 & $\mp$0.73 & $\mp 1.31$ \\

  1352Hz(1,3,1) & 0.27 & 0.63 & 1.00 & 1.37 & $-0.27$ & 0.10 & 0.46 & 0.83 \\

\hline

  PPH &

\multicolumn{4}{c} {$\omega=m\Omega+n^{1/2}\Omega_\bot$} &

\multicolumn{4}{c} {$\omega=m\Omega-n^{1/2}\Omega_\bot$}   \\

  $\omega_{\rm s}(m_{\rm s}, n_{\rm s},n)$  & 

  $m=0$&$m=1$&$m=2$&$m=3$&$m=0$&$m=1$ & $m=2$& $m=3$ \\

\hline

   133Hz(2,3,1) & 3.73 & 7.46 & 11.19 & 14.93 & $-3.73$ & 0.00 & 3.73 & 7.46 \\

   $-362$Hz(1,3,1) & $-1.37$ & $-2.73$ & $-4.10$ & $-5.46$ & 1.37 & 0.00 

                   & $-1.37$ & $-2.73$ \\

   628Hz(3,3,1) & 0.79 & 1.58 & 2.37 & 3.15 & $-0.79$ & 0.00 & 0.79 & 1.58 \\

   $\mp$857Hz(0,3,1) & $\mp$0.58 & $\mp$1.16 & $\mp$1.73 & $\mp 2.31$ 

      & $\pm 0.58$ & 0.00 & $\mp$0.58 & $\mp$1.16 \\

   1352Hz(1,3,1) & 0.37 & 0.73 & 1.10 & 1.46 & $-0.37$ & 0.00 & 0.37 & 0.73 \\

  495Hz(0 or 2,1,3) & 1.73 & 2.73 & 3.73 & 4.73 & $-1.73$ & $-0.73$

      & 0.27 & 1.27\\

  990Hz(1 or 3,1,3) & 0.87 & 1.37 & 1.87 & 2.37 & $-0.87$ & $-0.37$ & 0.13 & 0.63\\

   1485Hz(2 or 4,1,3) & 0.58 & 0.91 & 1.24 & 1.58 & $-0.58$ & $-0.24$ & 0.09 & 0.42 \\

   1980Hz(3,1,3) & 0.43 & 0.68 & 0.93 & 1.18 & $-0.43$ & $-0.18$ & 0.07 & 0.32 \\
\hline

\end{longtable}


\begin{longtable}{rl}

  \caption{Spin frequency and pairs of $\Delta\omega/\omega_{\rm s}=
        1$, 1/2, and 1/3, when resonances occur 
         at $\kappa=(1/2)\Omega (r=4.0r_{\rm g})$.\footnotemark}
 \label{table 7}

\endfirsthead







\hline\hline

 \multicolumn{1}{c} {$\omega_{\rm s}$} &

 \multicolumn{1}{c} {$\omega /\omega_{\rm s}$} \\

\hline

509Hz & (1.00, 2.00, 3.00, 4.00, 5.00, 6.00 ...)     \\

1018Hz& (0.50, 1.50, 2.50, 3.50, ...)                \\

1526Hz& [[[0.33, 0.67, 1.00, 1.33, 1.67, 2.00, 2.33, 2.67,...]]]  \\ 
2035Hz& [[0.25, 0.75, 1.25, 1.75, ....]]             \\

\hline
\end{longtable}
\footnotetext
  {Sets of oscillations of $\Delta\omega/\omega_{\rm s}=1$ are in parentheses,
   those of $\Delta\omega/\omega_{\rm s}=1/2$ are in double brackets, and
   those of $\Delta\omega/\omega_{\rm s}=1/3$ are triple brackets.}

\begin{longtable}{rl}

  \caption{Spin frequency and pairs of $\Delta\omega/\omega_{\rm s}=1$
           when resonances occur at $\kappa=(\sqrt{2}/2)\Omega
           (r=6.0r_{\rm g})$.}
 \label{table 8}
\endfirsthead

\hline\hline

 \multicolumn{1}{c} {$\omega_{\rm s}$} &

 \multicolumn{1}{c} {$\omega /\omega_{\rm s}$} \\

\hline

162Hz & (1.00, 2.00), (1.41, 2.41), (4.41, 5.41), (4.83, 5.83) \\

229Hz & (0.17, 1.17), (3.12, 4.12), (5.54, 6.54)               \\

325Hz & (1.21, 2.21), (2.90, 3.91)                             \\

392Hz & (0.83, 1.83), (1.00, 2.00), (2.24, 3.24), 
        (2.41, 3.41), (3.83, 4.83), (5.24, 6.24) \\

554Hz & (0.71, 1.71, 2.71, ...), (0.29, 1.29, 2.29,...)        \\

716Hz & (0.23, 1.23), (0.32, 1.32), (1.09, 2.09)                \\

879Hz & (0.45, 1.45)                           \\

946Hz & (0.34, 1.34), (0.41, 1.41), (1.00, 2.00), (1.59, 2.59)   \\ 

1270Hz& (0.18, 1.18), (0.62, 1.62)                             \\

1500Hz& (0.26, 1.26), (0.63, 1.63)             \\

2054Hz& (0.19, 1.19)                                           \\

\hline 

\end{longtable}

\begin{longtable}{rl}

  \caption{Spin frequency and pairs of $\Delta\omega/\omega_{\rm s}=1$
           and 1/2 when resonances occur at $\kappa=(\sqrt{3}/2)\Omega
           (r=12.0r_{\rm g})$.\footnotemark}
 \label{table 9}

\endfirsthead







\hline\hline

 \multicolumn{1}{c} {$\omega_{\rm s}$} &

 \multicolumn{1}{c} {$\omega /\omega_{\rm s}$} \\

\hline

52.5Hz& (3.23, 4.23), (6.96, 7.96)                    \\

143Hz & (0.18, 1.18), (1.55, 2.55), (2.92, 3.92)      \\

170Hz & (0.31, 1.31), (1.00, 2.00), (1.46, 2.46), (2.15, 3.15), 
        (3.31, 4.31), (4.46, 5.46) \\

220Hz & (0.65, 1.65), (1.53, 2.53), (2.41, 3.41)     \\

339Hz & (0.66, 1.66), (1.23, 2.23)                   \\

365Hz & (0.14, 1.14), (0.46, 1.46), (1.00, 2.00), (1.54, 2.54),

        [[1.54, 2.07]]$^*$     \\

418Hz & (0.34, 1.34), (0.81, 1.81), [[1.28, 1.75, 2.22]]$^*$                                \\

561Hz & [[0.26, 0.75]]$^*$, (0.30, 1.30), (0.65, 1.65)            \\

\hline
\end{longtable}
\footnotetext
  {Sets of oscillations of $\Delta\omega/\omega_{\rm s}=1$ are in parentheses
   and those of $\Delta\omega/\omega_{\rm s}\sim 1/2$ are double brackets with 
   superscript *}

\begin{longtable}{rl}

\caption{Spin frequency and pairs of $\Delta\omega/\omega_{\rm s}=1$
         when resonances occur at $\kappa=(\sqrt{3}-\sqrt{2})\Omega
        (r=3.34r_{\rm g})$.}
\label{table 10}
\endfirsthead

\hline\hline

 \multicolumn{1}{c} {$\omega_{\rm s}$} &

 \multicolumn{1}{c} {$\omega /\omega_{\rm s}$} \\

\hline
357Hz & (1.19, 2.19), (1.55, 2.55), (4.92, 5.92), (5.27, 6.27), 

        (9.01, 10.01)  \\

424Hz & (0.84, 1.84), (1.30, 2.30), (4.45, 5.45), (7.60, 8.60), 

        (10.74, 11.74), (13.89, 14.89) \\
552Hz & (0.65, 1.65), (0.77, 1.77), (3.06, 4.06), (3.18, 4.18), (5.60, 6.60),

        (8.01, 9.01)   \\

781Hz & (1.25, 2.25), (4.66, 5.66)                              \\

909Hz & (1.07, 2.07), (2.54, 3.54), (4.00, 5.00), (5.47, 6.47) \\

976Hz & (0.80, 1.80), (0.93, 1.93), (2.17, 3.17), (3.66, 4.66) \\

1757Hz& (0.20, 1.20), (0.31, 1.31), (1.07, 2.07), (1.83, 2.83) \\
1985Hz& (0.19, 1.19), (0.22, 1.22), (0.90, 1.90), (0.93, 1.93), (1.64, 2.64), 

        (2.35, 3.35)   \\

2113Hz& (1.09, 2.09)                                           \\  

2242Hz& (0.44, 1.44), (1.03, 2.03), (1.62, 2.62)               \\

2308Hz& (0.34, 1.34), (0.92, 1.92)                             \\

\hline
\end{longtable}

\begin{longtable}{rl}

\caption{Spin frequency and pairs of $\Delta\omega/\omega_{\rm s}= 1$ 
             and 1/2 when resonances occur at 
           $\kappa=(\sqrt{2}-1)\Omega (r=3.62r_{\rm g})$.\footnotemark}
\label{table 11}


\endfirsthead







\hline\hline

 \multicolumn{1}{c} {$\omega_{\rm s}$} &

 \multicolumn{1}{c} {$\omega /\omega_{\rm s}$} \\

\hline

489Hz & (2.41, 3.41), (3.83, 4.83, 5.83), (6.24, 7.24, 8.24), (9.66, 10.66) \\

691Hz & (0.71, 1.71, 2.71), (2.41, 3.41, 4.41), (4.12, 5.12), (5.83, 6.83) \\

1180Hz& (0.41, 1.41, 2.41, 3.41, 4.41), (0.59, 1.59)                        \\ 

1669Hz& (0.71, 1.71), (1.12, 2.12), (1.41, 2.41) \\ 

1871Hz& (0.26, 1.26), (0.63, 1.63), (0.89, 1.89), (1.52, 2.52)   \\

2360Hz& [[0.21, 0.71, 1.21, 1.71, 2.21]], [[0.29, 0.79]]               \\

2849Hz& [[0.17, 0.66]]$^*$, (0.66, 1.66), (0.41, 1.41)                               \\  
\hline
\end{longtable}
\footnotetext
  {Sets of oscillations of $\Delta\omega/\omega_{\rm s}=1$ are in parentheses,
   those of $\Delta\omega/\omega_{\rm s}=1/2$ are in double brackets, and
   those of $\Delta\omega/\omega_{\rm s}\sim 1/2$ are double brackets with 
   superscript *.}

\begin{longtable}{rl}

  \caption{Spin frequency and pairs of $\Delta\omega/\omega_{\rm s}=1$
           and 1/2 when resonances occur at $\kappa=(\sqrt{3}-1)\Omega
           (r=6.46r_{\rm g})$.\footnotemark}
 \label{table 12}

\endfirsthead







\hline\hline

 \multicolumn{1}{c} {$\omega_{\rm s}$} &

 \multicolumn{1}{c} {$\omega /\omega_{\rm s}$} \\

\hline

133Hz & (2.73, 3.73, 4.73, ...), (6.46, 7.46, 8.46),

            (10.19, 11.19), (13.93, 14.93)    \\

362Hz & (0.37, 1.37, 2.37, ...), (1.73, 2.73, 3.73, ...), 
        (3.10, 4.10, 5.10, ...)  \\

495Hz & (0.73, 1.73, 2.73, 3.73, 4.73, ... ), (0.27, 1.27, 2.27,...),

        [[1.73, 2.27]]$^*$  \\

628Hz & (0.58, 1.58), (0.79, 1.79), (1.37, 2.37), (2.15, 3.15)  \\

857Hz & (0.58, 1.58), (1.16, 2.15)      \\  

990Hz & [[0.37, 0.87, 1.37, 1.87, 2.37,...]], [[0.13, 0.63, ...]]   \\

1485Hz& [[0.09, 0.58]]$^*$, (0.24, 1.24), [[0.42, 0.91]]$^*$, 
        (0.58, 1.58)   \\

1980Hz& (0.18, 1.18), [[0.18, 0.68]], [[0.43, 0.93]]           \\

\hline
\end{longtable}
\footnotetext
  {Sets of oscillations of $\Delta\omega/\omega_{\rm s}=1$ are in parentheses,
   those of $\Delta\omega/\omega_{\rm s}=1/2$ are in double brackets, and
   those of $\Delta\omega/\omega_{\rm s}\sim 1/2$ are double brackets with 
   superscript *}

\begin{longtable}{rrrrcrr}

\caption{Observed kHz QPOs and their models.}
 \label{table 13}
\hline\hline

\endfirsthead

 &   \multicolumn{2}{c} {Observations} &
   \multicolumn{4}{c} {Models}  \\
\hline
Objects          & Spin    & Oscillations($\omega/\omega_{\rm s})$(Hz)
                 & Types, $r/r_{\rm g}$    & Spin($1.4M_\odot$) 
                 & \multicolumn{1}{c}{Oscillations} & $M/M_\odot$
                 \\
\hline
J1808.4$-$3658 & 400 Hz & 1.25(500),\ \  1.75(700) & GPH, 12.0 & 418 Hz 
                 & 1.28($m=1$), 1.75($m=2$) &    
                 \\
             &        & 1.03(410)                &  
             &    & 1.06($m=4$)            & 1.5
                 \\

\hline
KS 1731$-$260      & 524 Hz & 1.72(900), 2.21(1160)& GPH, 12.0 & 418 Hz
                 & 1.75($m=2$), 2.22($m=3$) & 
                 \\
                 &        & 2.30(1205)  & GGV, 12.0    & 
                 & 2.28($m=4$)            & 1.1
                 \\
\hline

4U 1702$-$43       & 330 Hz & 1.98(655), 3.03(1000) & GPH, 12.0 & 365 Hz  
                 & 2.00($m=2$), 3.07($m=4$) & 
                 \\
                 &        & 2.12(700), 2.73(\, 902) & GGV, 12.0 &  
                 & 2.07($m=3$), 2.61($m=4$)  & 1.6
                 \\       
                 &        &                      & GPH/GGV, \ 4.0 & 
                   509 Hz & 2.00($m=0$), 3.00($m=1$) & 2.2
                 \\
\hline

4U 1728$-$34       & 364 Hz & 1.40(510), 2.40($\,$ 875) & GPH/GGV, \ 6.0 &
                   162 Hz & 1.41($m=1$), 2.41($m=0$) & 0.62
                 \\
\hline

4U 1636$-$53       & 582 Hz & 1.55(900), 2.04(1190)& GGV, 12.0 & 365 Hz
                 & 1.54($m=2$), 2.07($m=3$) & 0.88
                 \\
\hline

Aq1X-1           & 549 Hz & 1.22(670), 1.69(\, 930) & PGH, 12.0 & 339 Hz
                 & 1.23($m=3$), 1.66($m=2$) & 
                 \\

                 &        & 1.89(1040) &    &    &  1.81($m=4$)

                 & 0.86
                 \\ 
\hline
\end{longtable}

\end{document}